\documentclass[12pt]{iopart}

\usepackage{bm}
\usepackage{epsfig}
\usepackage{citesort}
\usepackage{graphicx}
\eqnobysec

\newcommand{\lwig}{\mbox{\;\raisebox{.3ex}
    {$<$}$\!\!\!\!\!$\raisebox{-.9ex}{$\sim$}\;}}
\newcommand{\gwig}{\mbox{\;\raisebox{.3ex}
    {$>$}$\!\!\!\!\!$\raisebox{-.9ex}{$\sim$}}\;}
\newcommand{\lambdabar}{{\hbox{$\lambda$\kern-1.ex\raise+0.45ex\hbox{--}}}}

\begin{document}

\begin{flushright}
{\large \tt LAPTH-1242/08 \\
MPP-2008-33}
\end{flushright}

\title[Nonlinear corrections to the cosmological matter power spectrum]%
{Nonlinear corrections to the cosmological matter power spectrum and scale-dependent galaxy bias: implications for parameter estimation}

\author{Jan~Hamann}
\address{LAPTH (Laboratoire d'Annecy-le-Vieux de Physique
Th\'eorique, CNRS UMR5108 \&~Universit\'e de Savoie), BP 110, F-74941
 Annecy-le-Vieux Cedex, France}

\author{Steen~Hannestad}
\address{Department of Physics and Astronomy, University of Aarhus \\
Ny Munkegade, DK-8000 Aarhus C, Denmark}

\author{Alessandro Melchiorri}
\address{Physics Department and sezione INFN, University of Rome ``La Sapienza''\\
Ple Aldo Moro 2, I-00185 Rome, Italy}

\author{Yvonne~Y.~Y.~Wong}
\address{Max-Planck-Institut f\"ur Physik (Werner-Heisenberg-Institut) \\
F\"ohringer Ring 6, D-80805 M\"unchen, Germany}

\ead{\mailto{hamann@lapp.in2p3.fr}, \mailto{sth@phys.au.dk}, \mailto{alessandro.melchiorri@roma1.infn.it},
 \mailto{ywong@mppmu.mpg.de}}

\begin{abstract}
We explore and compare the performances of two nonlinear correction
and scale-dependent biasing models for the extraction of cosmological
information from galaxy power spectrum data, especially in the context
of beyond-$\Lambda$CDM cosmologies.  The first model is the well known
$Q$~model, first applied in the analysis of 2dFGRS data.  The second,
the $P$~model, is inspired by the halo model, in which nonlinear
evolution and scale-dependent biasing are encapsulated in a single
non-Poisson shot noise term.  We find that while both models perform
equally well in providing adequate correction for a range of galaxy
clustering data in standard $\Lambda$CDM cosmology and in extensions
with massive neutrinos, the $Q$~model can give unphysical results
in cosmologies containing a
subdominant free-streaming dark matter whose temperature depends on
the particle mass, e.g., relic thermal axions, unless a suitable prior
is imposed on the correction parameter.  This last case also exposes
the danger of analytic marginalisation, a technique sometimes used in
the marginalisation of nuisance parameters.  In contrast, the
$P$~model suffers no undesirable effects, and is the recommended
nonlinear correction model also because of its physical transparency.

\end{abstract}
\maketitle

\section{Introduction}

The past decade saw an explosion of precision cosmological
measurements.  Foremost amongst these is the observation of
temperature and polarisation fluctuations in the cosmic microwave
background (CMB) by a range of
experiments~\cite{Hinshaw:2008kr,Nolta:2008ih,Reichardt:2008ay,Jones:2005yb,Piacentini:2005yq,Montroy:2005yx}.
The
distribution of large-scale structures (LSS) has also been mapped to
unprecedented breadths and depths by galaxy redshift surveys such as
the Two-Degree Field Galaxy Redshift Survey
(2dFGRS)~\cite{Cole:2005sx} and the Sloan Digital Sky Survey
(SDSS)~\cite{Tegmark:2006az,Percival:2006gt}.  Together with
observations of distant type Ia
supernovae~\cite{Riess:1998cb,Perlmutter:1998np}, these measurements
have fostered the emergence of a benchmark framework---the adiabatic,
nearly scale-invariant, ``vanilla'' $\Lambda$CDM model---based on which
one can test for evidence of new physics.

Clustering statistics of galaxies as probed by surveys like 2dF and
SDSS are particularly well suited for the exploration of physics that
introduce new effects on length scales ${\cal O}(10) \to {\cal
O}(100) \ h^{-1} \ {\rm Mpc}$.  A classic example is the possibility to detect a
subdominant component of free-streaming hot dark matter
(HDM)~\cite{Hu:1997mj}, notably massive
neutrinos~\cite{Hannestad:2006zg,Lesgourgues:2006nd} and variants such
as thermal
axions~\cite{Hannestad:2003ye,Hannestad:2005df,Hannestad:2007dd,Hannestad:2008js,Melchiorri:2007cd},
light gravitinos~\cite{Viel:2005qj}, {\it et~cetera}.  The sensitivity
of these surveys at small length scales also lends a greater lever arm
to the search for features in the primordial density perturbation
power spectrum, possible remnants of inflationary
physics~\cite{Covi:2006ci,Hamann:2007pa}.  Last but not least, since
the power spectrum of large-scale structures probes uniquely the
parameter combination~$\Omega_m h$, it helps to lift the degeneracy
between---and hence tighten the constraints on---the physical matter
density~$\Omega_m h^2$ and the Hubble parameter~$h$ from CMB
observations alone.

Usage of data from galaxy clustering surveys is based on the premise
that one can reliably predict the distribution of galaxies, at least
on a statistical basis, from theory.  This is complicated by a number
of factors.  First, galaxies are necessarily collapsed objects, i.e.,
they have undergone a phase of nonlinear evolution.  Using them as
tracers of the underlying matter field implicitly assumes we know how
to relate the two distributions to one another.  A reasonable
assumption is that on sufficiently large scales the power spectra of
galaxies and of the matter field are identical up to a constant
normalisation, or bias, factor.  But this bias relation is expected to
become scale-dependent when the dimesionless power spectrum of the
galaxies $\Delta_{\rm gal} \equiv k^3 P_{\rm gal}/2 \pi^2$ exceeds
unity~\cite{Benson:1999mv,Blanton:1999gd}.  
Indeed, the
apparent tension between the 2dF and the SDSS galaxy power spectra is
now believed to have originated from a more strongly scale-dependent
bias factor for the red galaxies dominating the SDSS galaxy
catalogue~\cite{Cole:2006kn,Sanchez:2007rc}.
On the theoretical front,
a good deal of recent effort has also been devoted to understanding 
the origin of scale-dependent biasing 
(e.g., \cite{McDonald:2006mx,Smith:2006ne}).

Second, galaxy positions are inferred from their redshifts.  However,
the peculiar motions of the galaxies, particularly when amplified by
virialisation, induce additional Doppler shifts that can potentially
obscure the inference.  This is known as redshift space distortion,
and on small length scales requires corrections beyond linear
perturbation theory.  Third, the clustering of the underlying dark
matter field itself becomes nonlinear on scales~$k \gwig 0.2 \ h \
{\rm Mpc}^{-1}$.  Thus, how reliably one can extract cosmological
information from galaxy clustering statistics depends crucially on how
well one can model these three nonlinear effects.

While all three effects can in principle be modelled by numerical
simulations, these simulations, and indeed our understanding of galaxy
formation, are not yet at a stage where one can reliably predict the
power spectrum of galaxies as a function of galaxy type and redshift
given some underlying cosmological model.  In the meantime, nonlinear
evolution and scale-dependent biasing must be modelled empirically as
a systematic effect and the associated nuisance parameters
marginalised when extracting cosmological information from galaxy
clustering surveys, especially in beyond-$\Lambda$CDM cosmologies.

In this connection, Cole {\it et al.}~\cite{Cole:2005sx} recently
proposed a correction formula which maps directly between the matter
power spectrum calculated from linear perturbation theory~$P_{\rm
lin}(k)$ and the power spectrum expected for the galaxies~$P_{\rm
gal}(k)$,
\begin{equation}
\label{eq:Qmodel}
P_{\rm gal}(k) = b^2 \frac{1+Q_{\rm nl} k^2}{1+A_{\rm nl} k} P_{\rm lin}(k).
\end{equation}
The formula is partially calibrated against $\Lambda$CDM-based
semi-analytic galaxy formation simulations ($A_{\rm nl}=1.4$ for redshift space, 
and $1.7$ for real space), and
contains two free parameters ($b$ and $Q_{\rm nl}$) to be fixed by
observational data.  Equation~(\ref{eq:Qmodel}) has been applied
to the galaxy power spectra of 2dF~\cite{Cole:2005sx}
and the SDSS luminous red galaxy (LRG) sample~\cite{Tegmark:2006az} to
test standard vanilla cosmology.  However, there is {\it a priori} no
guarantee that its usefulness extends also to cosmologies beyond
$\Lambda$CDM.  Indeed, as we shall show below,
equation~(\ref{eq:Qmodel}) can be highly pathological when applied to
certain classes of cosmological models containing a subdominant
component of free-streaming dark matter.

In comparison, a conceptually more appealing framework in which to
discuss nonlinear corrections is the halo
model~\cite{Seljak:2000gq,Peacock:2000qk,Ma:2000ik,Cooray:2002di}.
Building on the assumptions of (i)~hierarchical clustering, and
(ii)~that galaxies form only inside dark matter halos, halo
model-based nonlinear corrections can in principle be made applicable
to all hierarchical CDM cosmologies.  The minimal model proposed in
references~\cite{Seljak:2000jg,Schulz:2005kj,Guzik:2006bu}, for
example,
\begin{equation}
\label{eq:Pmodel}
P_{\rm gal}(k)=b^2 P_{\rm lin}(k) + P_{\rm shot},
\end{equation}
where~$b$ and~$P_{\rm shot}$ are free parameters, does not demand
vanilla $\Lambda$CDM as the sole input cosmology.

In the present work, we explore and compare the performance of these
nonlinear correction models in some detail.  We confront them with
various observed galaxy power spectra, especially in the context of
beyond-$\Lambda$CDM cosmologies.  We discuss parameter degeneracies and
the role of priors on the
nuisance parameters.

The paper is organised as follows.  We describe first in
section~\ref{sec:data} the galaxy clustering data sets used in the
analysis. Section \ref{sec:bias} contains a more detailed discussion
of the two nonlinear models we wish to explore. In sections
\ref{sec:vanilla}, \ref{sec:neutrinos} and \ref{sec:axions} we test
the nonlinear models against data for standard vanilla cosmology,
vanilla with massive neutrinos, and vanilla with thermal axions
respectively.  We conclude in section \ref{sec:conclusions}.

\section{Data sets \label{sec:data}}

We use the publicly available galaxy power spectra from the following
galaxy catalogues.

\paragraph{2dF} This data set comes from the final data release of the
Two-Degree Field Galaxy Redshift Survey~\cite{Cole:2005sx}.  We use up
to 36 data bands, corresponding to redshift space power spectrum data 
for wavenumbers
$0.02 \lwig k/h \ {\rm Mpc}^{-1} \lwig 0.18$.

\paragraph{SDSS-2~main} Real space power spectrum of the
main galaxy sample from the Sloan Digital Sky Survey data release
2~\cite{Tegmark:2003ud}.  We use up to 19 data bands, i.e., $0.016
\lwig k/h \ {\rm Mpc}^{-1} \lwig 0.2$.

\paragraph{SDSS-4~LRG}  Real space power spectrum of the luminous red galaxies from
the Sloan Digital Sky Survey data release 4~\cite{Tegmark:2006az}.
The 20 data bands correspond to wavenumbers
$0.012 \lwig k/h \ {\rm Mpc}^{-1} \lwig 0.2$.

\paragraph{WMAP-3} We also use at times CMB data from the Wilkinson
Microwave Anisotropy Probe experiment after three years of
observations~\cite{Hinshaw:2006ia,Page:2006hz,Spergel:2006hy}, mainly
for the construction of priors on certain cosmological
parameters. This calculation is performed using version~2 of the
likelihood package provided by the WMAP team on the LAMBDA web
page~\cite{bib:lambda}.

\section{Two nonlinear  models \label{sec:bias}}

\subsection{The $Q$ model}

We refer to the correction formula~(\ref{eq:Qmodel}) as the $Q$~model.
For $\Lambda$CDM cosmologies, galaxy formation simulations suggest
that the parameter~$A_{\rm nl}$ can be held fixed at $A_{\rm
nl}=1.4$ in redshift space and at $A_{\rm nl}=1.7$ in real space~\cite{Cole:2005sx}.  
The parameter~$Q_{\rm nl}$, however,
exhibits a strong dependence on the galaxy type.  Fitting the
$Q$~model to the 2dF galaxy power spectrum at $0.02 \lwig k/h \ {\rm
Mpc}^{-1} \lwig 0.3$, Cole {\it et al.} found $Q_{\rm nl}=4.6 \pm 1.5$
for vanilla cosmology~\cite{Cole:2005sx}.  Tegmark {\it et
al.}~\cite{Tegmark:2006az} applied the same model to SDSS-4~LRG, for
which $Q_{\rm nl}= 30 \pm 4$ provides a good fit.%
\footnote{Tegmark {\it et al.}~\cite{Tegmark:2006az} used $A_{\rm nl} = 1.4$,
although strictly speaking $A_{\rm nl} = 1.7$ is the more appropriate value
for the real space power spectrum of SDSS-4~LRG.  However, 
at the present level of precision,  our tests show that
adopting the correct $A_{\rm nl}=1.7$  only causes a statistically insignificant upward 
shift in the best-fit $Q_{\rm nl}$ ($\Delta Q_{\rm nl} \sim 2$), while the estimates 
of other cosmological parameters remain unaffected.
Henceforth we shall use exclusively $A_{\rm nl}=1.4$ for
both real and redshift space power spectra.}

Note that the case of $Q_{\rm nl}=0$ is not equivalent to
no nonlinear correction, since a nonzero~$A_{\rm nl}$ parameter
also modulates the power spectrum in a scale-dependent way.
For $Q_{\rm nl} \lwig 7$, the $Q$~model suppresses the power spectrum
at $k \lwig 0.2 \ h \ {\rm Mpc}^{-1}$. However, for $Q_{\rm nl}$~values 
as large as $30$ such as required by the SDSS-4 LRG, the main role of 
the $Q$~model is to add power at $k \gwig 0.07 \ h \ {\rm Mpc}^{-1}$.

\subsection{The $P$ model}

The functional form of the $Q$~model~(\ref{eq:Qmodel}) was recently
criticised in reference~\cite{Smith:2006ne} for its incorrect
dependence on~$k$.  Specifically, it lacks a constant term to account
for the presence of non-Poisson shot noise, a generic consequence of
the assumption that galaxies form exclusively in halos.  Indeed, from
the halo model
\cite{Seljak:2000gq,Peacock:2000qk,Ma:2000ik,Cooray:2002di}, one
should expect a correction formula with the skeletal form,
\begin{equation}
P_{\rm gal}(k)=P_{\rm 2h}(k) + P_{\rm 1h}(k).
\end{equation}
Here, the two-halo term, $P_{\rm 2h}(k) = b^2(k) P_{\rm lin}(k)$, 
arises from correlations between galaxies in two different halos, and 
approximates the familiar linear bias relation on large scales,
\begin{equation}
\label{eq:2halo}
P_{\rm 2h}(k) \approx  b^2 P_{\rm lin}(k).
\end{equation}
The one-halo term
\begin{equation}
\label{eq:1halo}
P_{\rm 1h}(k) \approx {\rm const.} \equiv P_{\rm shot}
\end{equation}
accounts for correlations within the same halo and is approximately
independent of the exact spatial distribution of the galaxies within a
halo provided $k$ is not too large.  Combining equations~(\ref{eq:2halo}) and 
(\ref{eq:1halo}) we recover the minimal model~(\ref{eq:Pmodel}).

The role of the shot noise term $P_{\rm shot}$ is to add power at small length scales, so 
that the ratio $P_{\rm gal}(k)/P_{\rm lin}(k)$ is effectively scale-dependent
at large $k$ values.
This is in contrast with the $Q$~model~(\ref{eq:Qmodel}), which for some values of the 
nonlinear parameter~$Q_{\rm nl}$ suppresses the power spectrum on the observable scales.
Interestingly, taken at face value, the $P$~model also predicts a significant $k$-dependence 
for $P_{\rm gal}(k)/P_{\rm lin}(k)$ as $k \to 0$
when $P_{\rm shot}$ once again rises above \mbox{$b^2 P_{\rm lin}(k) \sim k$}~\cite{Seljak:2000gq}.
It should be noted however that
this small-$k$ behaviour is not present for 
dark matter clustering, since momentum conservation demands that
the dark matter 
power spectrum falls off faster than $P(k) \propto k^4$ as $k\to0$~\cite{Zeldovich:1969sb}, 
a behaviour also observed in numerical simulations~\cite{Smith:2002dz,Crocce:2007dt}.
On the other hand, a non-vanishing shot noise on large scales is in principle not forbidden for galaxy clustering, 
since tracers need not conserve momentum.   Whether or not this is  so remains to be understood.

In the present work we take the view that since galaxy clustering has not
been observed on scales where the shot noise behaviour may
become problematic (\mbox{$k \ll 0.01 \ h \ {\rm Mpc}^{-1}$}), equation~(\ref{eq:Pmodel}) 
constitutes a sufficient phenomenological model to describe
the galaxy power spectrum at $k \gwig 0.01 \ h \ {\rm Mpc}^{-1}$.

Additional modifications to the basic $P$~model~(\ref{eq:Pmodel}) 
to account for the damping of baryon acoustic
oscillations and other nonlinear mode coupling effects have been
discussed in the literature~\cite{Smith:2006ne,Crocce:2007dt,Smith:2007gi,Huff:2006gs}.
These generally lead to nonlinear models of the form
\begin{equation}
P_{\rm gal}(k) = b^2 A(k) P_{\rm lin}(k) + P_{\rm shot},
\end{equation} 
where $A(k)$ is some function of $k$ that depends also on the galaxy type.
It is also possible to extend 
the halo model to include redshift space 
distortion~\cite{Seljak:2000jg,Huff:2006gs,White:2000te}. 
We ignore these additional corrections in the present analysis in order to keep the number of extra fit 
parameters to a minimum.  However, we emphasise that 
the effects encapsulated by $A(k)$ 
will become increasingly important as more precise data from future galaxy redshift 
surveys become available.

\section{Test 1: vanilla \label{sec:vanilla}}

\subsection{Set-up}

To compare the performance of the two nonlinear models, we test them
against galaxy clustering data in three minimal parameter spaces:
\begin{enumerate}
\item  $\Omega_m h,\Omega_b h^2,h,n_s,\ln (10^{10} {\cal A})$,
\item $\Omega_m h, \Omega_b h^2,h,n_s,\ln (10^{10} {\cal A}), P_{\rm shot}$, and
\item $\Omega_m h,\Omega_b h^2,h,n_s,\ln (10^{10} {\cal A}),Q_{\rm nl}$.
\end{enumerate}
We take the geometry of the universe to be flat, and the initial
conditions adiabatic.  The parameter ${\cal A} \equiv b^2 A_s$
accounts for the normalisation of the galaxy power spectrum, and
incorporates both the amplitude of the primordial scalar
perturbations~$A_s$ and the constant galaxy bias factor~$b$.  We do
not use any nonlinear correction for parameter space~(i), i.e.,
$P_{\rm gal}(k) = b^2 P_{\rm lin}(k)$.  For parameter spaces~(ii) and
(iii), we use the $P$~model~(\ref{eq:Pmodel}) and the
$Q$~model~(\ref{eq:Qmodel}) respectively.

Note the definition of the matter density parameter~$\Omega_m h$.  We
choose this parameterisation because the turning point of the matter
power spectrum is sensitive to the comoving Hubble radius at
matter--radiation equality, which, for fixed values of $h$, depends
on~$\Omega_m h$, not the physical matter density~$\Omega_m h^2$.  For
fixed values of $n_s$ and~$h$, $\Omega_m h$ alone determines the broad
shape of the matter power spectrum within the $\Lambda$CDM framework.

We use standard Bayesian inference techniques and the Markov Chain
Monte Carlo package {\sc CosmoMC} \cite{Lewis:2002ah,cosmomc} to
explore the posterior hypersurfaces as functions of the model
parameters and the galaxy clustering data sets of
section~\ref{sec:data}.  Here, we note that within the vanilla
framework, the physical baryon density~$\Omega_b h^2$, the Hubble
parameter~$h$, and the scalar spectral index~$n_s$ can be individually
well constrained by CMB observations.  This information is
encapsulated in a set of ``WMAP-3 priors'' in table~\ref{tab:priors},
which we apply when varying these three parameters.  This approach
differs slightly from the more common practice of fixing the parameter
values $h=0.72$ and $n_s=1$ adopted in, e.g.,
references~\cite{Cole:2005sx,Sanchez:2007rc}.
Reference~\cite{Tegmark:2006az} further fixes $\Omega_b h^2 = 0.0223$.
Our approach has the advantage that it properly takes into account the
uncertainties on these parameters and thus avoids two inherent dangers of
fixed parameter analyses: biased parameter estimates and underestimated
errors.  Additionally, it permits a consistent comparison of our
$\Omega_m h$ constraints not only between different galaxy clustering
data sets, but also with those obtained from CMB observations alone.

%%%%%%%%%%%%%%%%%%%%%%%%%%%%%%%%%%%%%%%%%%%%%%%%
%%%%%%%%%%%%%%%%%%%%%%%%%%%%%%%%%%%%%%%%%%%%%%%%
\begin{table}
{\footnotesize
\caption{Priors used in our vanilla and vanilla+massive neutrinos analyses.
{\it Top}:~WMAP priors based on the three-year data release.  These
are approximated as perfect Gaussians, and we give here their
respective mean and standard deviation.  {\it Bottom}: Top-hat priors
for the remaining parameters.  \label{tab:priors}} \hskip25mm
\begin{tabular}{lcc}
 \br
 Parameter & vanilla & vanilla+neutrinos\\
 \mr
WMAP-3 priors \\
 $\Omega_{b} h^2$   &   $0.02229 \pm 0.00073$ & $0.02158 \pm 0.00082$\\
 $h$    & $0.732 \pm 0.032$  & $0.651 \pm 0.054$         \\
 $n_s$   & $0.958 \pm 0.016$  & $0.941 \pm 0.022$        \\
 \mr
Top-hat priors\\
$\Omega_m h$ & \centre{2}{0.05--0.72} \\
 $\ln (10^{10} {\cal A})$ & \centre{2}{1--5} \\
 $P_{\rm shot}$ & \centre{2}{0--${\rm min}(P_{\rm obs})$}\\
 $Q_{\rm nl}$ & \centre{2}{0--50} \\
\br
\end{tabular}
}
\end{table}
%%%%%%%%%%%%%%%%%%%%%%%%%%%%%%%%%%%%%%%%%%%%%%%%%%%
%%%%%%%%%%%%%%%%%%%%%%%%%%%%%%%%%%%%%%%%%%%%%%%%%%%

Broad, top-hat priors are imposed on the remaining parameters
(table~\ref{tab:priors}).  For the parameter~$Q_{\rm nl}$, we choose
the range 0--50 for all three data sets.  In the absence of additional
information from, e.g., galaxy formation simulations, the upper limit
of this prior is somewhat arbitrary.  We will discuss this point in
more detail in section~{\ref{sec:axions}}.  For the $P_{\rm shot}$
prior, the upper limit ${\rm min}(P_{\rm obs})$ denotes the minimum
clustering power measured by a survey.  In other words, the linear
matter power spectrum $P_{\rm lin}(k)$ must not be negative anywhere.
For 2dF, SDSS-2~main, and SDSS-4~LRG, ${\rm min}(P_{\rm
obs})=\{4000,1200,9300\}$, respectively.

\subsection{The internal test: do the individual data sets call for nonlinear correction? \label{sec:results}}

\begin{figure}[t]
%\hspace{10mm}
\includegraphics[width=15.5cm]{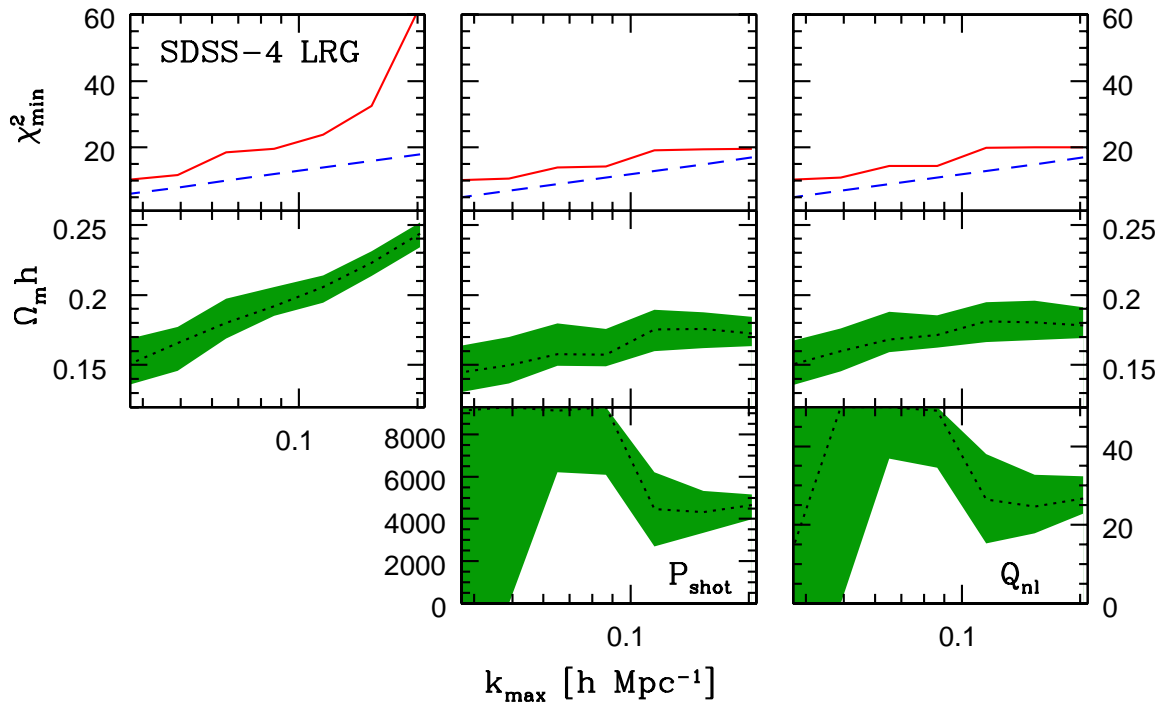}
\caption{1D marginal 68\% MCIs for $\Omega_m h$ and $1\sigma$ intervals
(see main text for definitions)
for the nonlinear parameters
$P_{\rm shot}$ and $Q_{\rm nl}$ as functions of $k_{\rm max}$ using
the SDSS-4~LRG power spectrum (green/shaded regions).  {\it Left}: No
nonlinear correction. {\it Centre}: Correction with the $P$~model.
{\it Right}: Correction with the $Q$~model.  In each case the black
dotted line indicates the 1D mode.  The top plot in each column shows
the minimum $\chi^2$ (red/solid) and the number of degrees of freedom
in the fit (blue/dash).\label{fig:LRG}}
\end{figure}

\begin{figure}[t]
%\hspace{10mm}
\includegraphics[width=15.5cm]{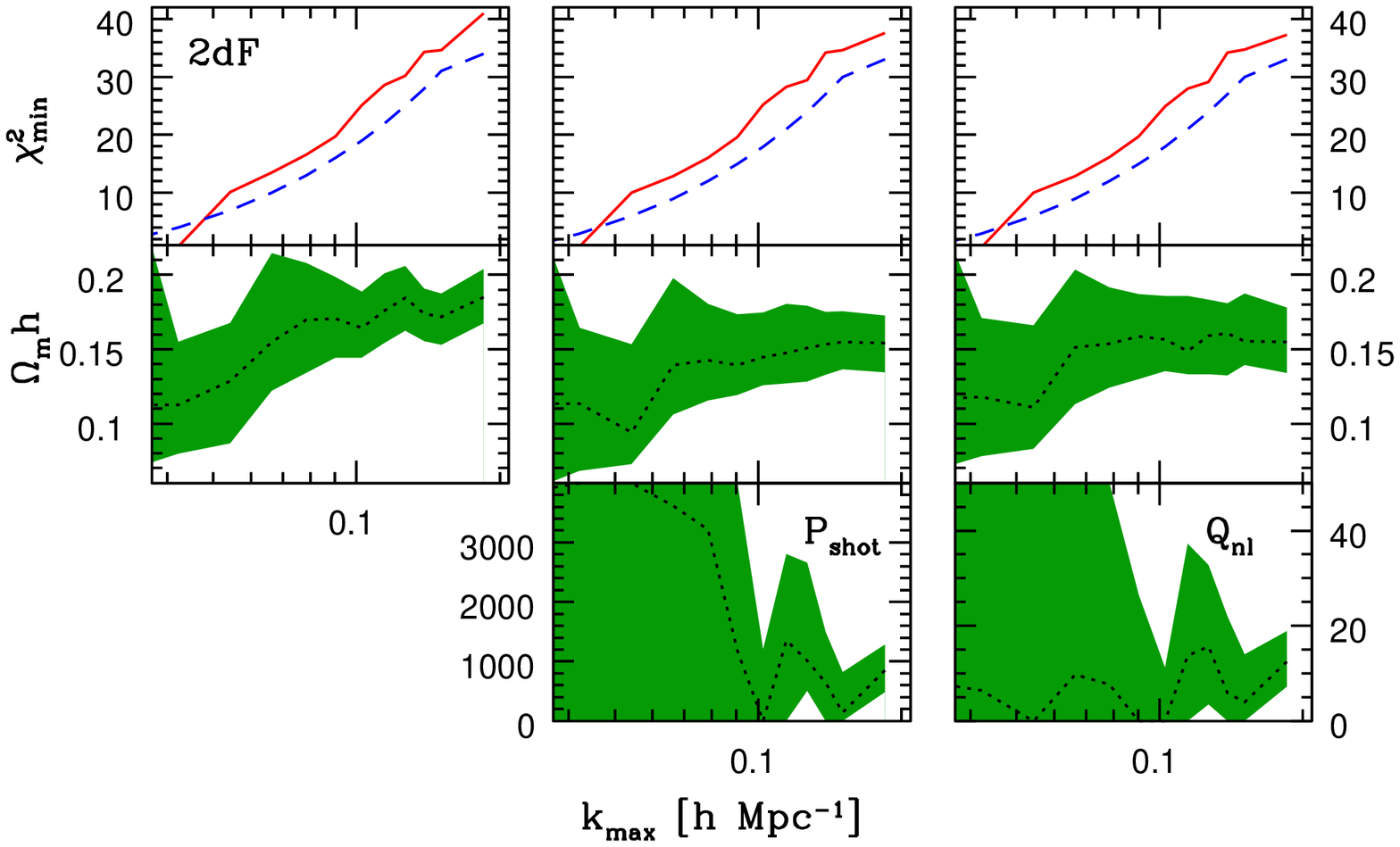}
\caption{Same as figure~\ref{fig:LRG}, but for
2dF.\label{fig:2dF}}
%\end{figure}
%%%%%%%%%%%%%%%%%%%%%%%%%%%%%%%%%%%%%%%%%%%%%%%%%%%%%%%%%%
%%%%%%%%%%%%%%%%%%%%%%%%%%%%%%%%%%%%%%%%%%%%%%%%%%%%%%%%%%
%
%%%%%%%%%%%%%%%%%%%%%%%%%%%%%%%%%%%%%%%%%%%%%%%%%%%%%%%%%%
%%%%%%%%%%%%%%%%%%%%%%%%%%%%%%%%%%%%%%%%%%%%%%%%%%%%%%%%%%
%\begin{figure}[t]
%\hspace{10mm}
\vspace{12mm}
\includegraphics[width=15.5cm]{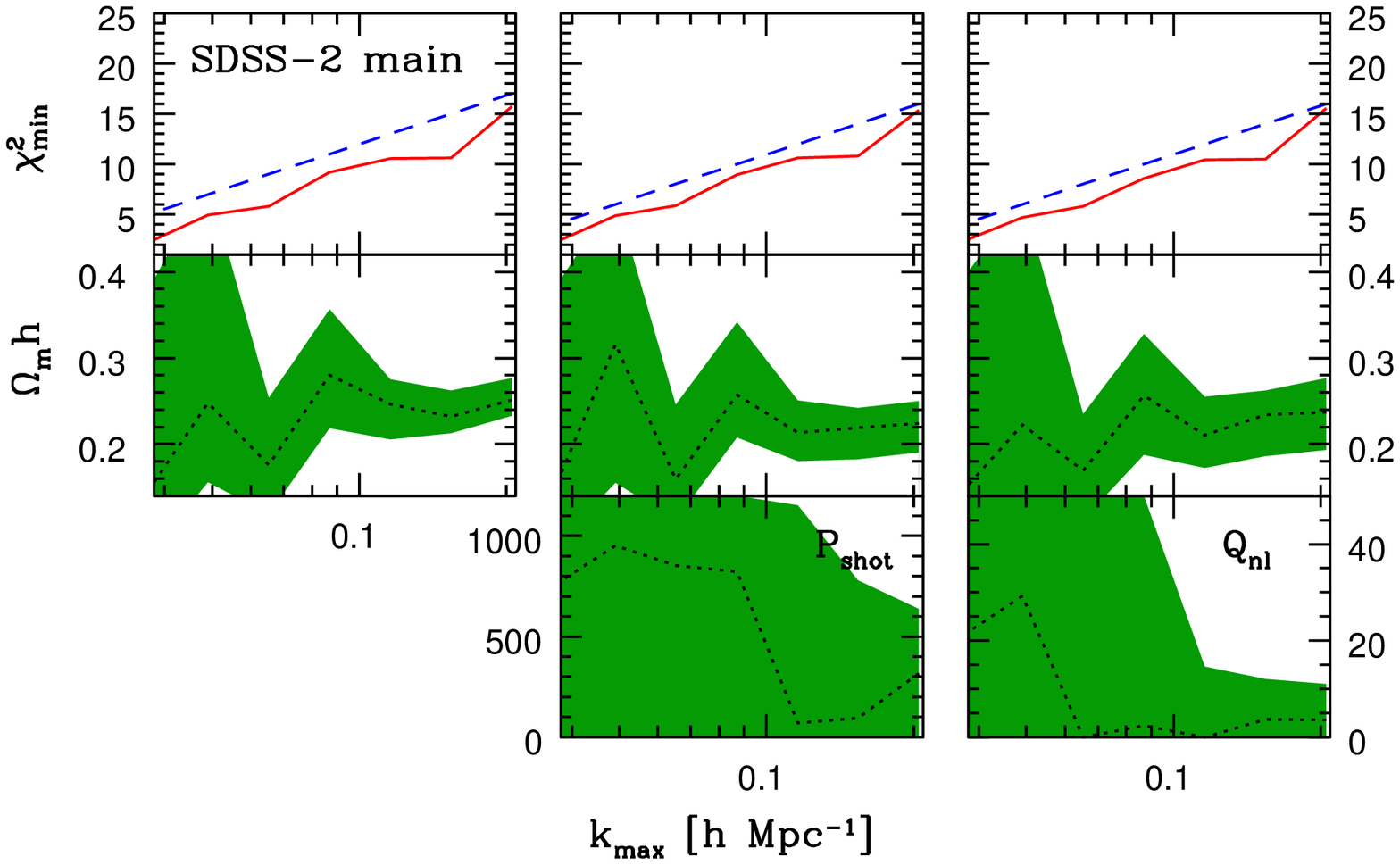}
\caption{Same as figure~\ref{fig:LRG}, but for
SDSS-2~main.\label{fig:main}}
\end{figure}

Figures~\ref{fig:LRG} to \ref{fig:main} show the 1D marginal
constraints on the parameters
\mbox{$\{\Omega_m h,P_{\rm shot},Q_{\rm nl}\}$} and the corresponding
minimum $\chi^2$ and numbers of degrees of freedom (d.o.f.)
as functions of the maximum wavenumber $k_{\rm max}$
included in the analysis.  The number of d.o.f.\ is defined here
as the number of data points plus the number of priors, minus the number of fit
parameters;  its actual value might be slightly lower, because the power
spectrum data
points are not completely uncorrelated owing to overlapping window functions.

%%%%%%%%%%%%%%%%%%%%%%%%%%%%%%%%%%%%%%%%%%%%%%%%%%%%%%%%%%
%%%%%%%%%%%%%%%%%%%%%%%%%%%%%%%%%%%%%%%%%%%%%%%%%%%%%%%%%%
%%%%%%%%%%%%%%%%%%%%%%%%%%%%%%%%%%%%%%%%%%%%%%%%%%%%%%%%%%
%%%%%%%%%%%%%%%%%%%%%%%%%%%%%%%%%%%%%%%%%%%%%%%%%%%%%%%%%%

%%%%%%%%%%%%%%%%%%%%%%%%%%%%%%%%%%%%%%%%%%%%%%%%%%%%%%%%%%
%%%%%%%%%%%%%%%%%%%%%%%%%%%%%%%%%%%%%%%%%%%%%%%%%%%%%%%%%%
%%%%%%%%%%%%%%%%%%%%%%%%%%%%%%%%%%%%%%%%%%%%%%%%%%%%%%%%%%%
%%%%%%%%%%%%%%%%%%%%%%%%%%%%%%%%%%%%%%%%%%%%%%%%%%%%%%%%%%%

For $\Omega_m h$, the green/shaded regions correspond to the
1D marginal 68\% minimum credible intervals (MCI), while
the 1D
modes are indicated by black/dotted lines.
On the other hand,
the marginalised posterior distributions in $P_{\rm shot}$ and $Q_{\rm
nl}$ are often extremely flat, especially  at small values of $k_{\rm max}$.
While it is technically possible to construct
credible intervals in these cases, quoting such an interval would
distract from the fact that the parameters are essentially
unconstrained. Therefore, instead of Bayesian intervals, we give
in figures~\ref{fig:LRG} to \ref{fig:main} for $P_{\rm shot}$ and $Q_{\rm nl}$
the parameter regions satisfying
\begin{equation}
- 2 \, (\ln \mathcal{P} - \ln \mathcal{P}_{\rm max}) < 1,
\end{equation}
where $\mathcal{P}$ is the 1D marginal posterior and $\mathcal{P}_{\rm max}$ denotes its
value at the 1D mode.  We loosely label this the ``$1 \sigma$'' interval.
For large $k_{\rm max}$ values, especially $k_{\rm max} \sim 0.2 \ h \ {\rm Mpc}^{-1}$,
 where the posterior distributions are approximately
Gaussian, this $1 \sigma$ interval is identical to the 68\% MCIs.
See
reference~\cite{Hamann:2007pi} for more detailed discussions of the various
statistical quantities.

\paragraph{Changes in the $\Omega_m h$ estimates}

The effects of nonlinear correction are most evident when we include
data beyond $k_{\rm max} \sim 0.1 \ h \ {\rm Mpc}^{-1}$.  Here,
nonlinear correction generally shifts the $\Omega_m h$ estimates to
lower values relative to the case with no correction, irrespective of
the nonlinear model used.  For SDSS-4~LRG, this shift is very
dramatic: at $k_{\rm max} \sim 0.2 \ h \ {\rm Mpc}^{-1}$ the 68\% MCI
moves down by an amount comparable to six or seven times its
half-width (figure~\ref{fig:LRG}).  In contrast, the shifts induced
for 2dF and \mbox{SDSS-2 main} by either nonlinear model are mild: at $k_{\rm
max} \sim 0.2 \ h \ {\rm Mpc}^{-1}$ a small overlap between the 68\%
MCIs before and after correction can still be seen in
figures~\ref{fig:2dF} and \ref{fig:main}.

\paragraph{The $\chi^2$ test}
The need for nonlinear correction in the case of SDSS-4~LRG is
corroborated by a comparison of the minimum $\chi^2$ values.  Between
correction and no correction, figure~\ref{fig:LRG} shows that at $k_{\rm max} \sim 0.2 \ h \
{\rm Mpc}^{-1}$,
$\chi^2_{\rm min}$ changes from $\sim 60$ for 18~d.o.f.\ to 
$\sim 20$ for 17~d.o.f.,
again irrespective of
the nonlinear model used.
  On the other hand, similar comparisons for 2dF and
for SDSS-2~main in figures~\ref{fig:2dF} and \ref{fig:main} do not
indicate any urgent need for an extra correction parameter: for 2dF,
$\Delta \chi^2_{\rm min} \sim 4$; for SDSS-2~main, $\Delta \chi^2_{\rm
min}$ is virtually negligible.

\paragraph{Internal consistency}  We consider a data set internally consistent if
the $\Omega_m h$ credible intervals agree for all choices of $k_{\rm
max}$.  For SDSS-4~LRG, figure~\ref{fig:LRG} clearly demonstrates that
nonlinear correction is necessary in order to achieve some semblance
of internal consistency.  A small discrepancy remains after
correction: at $k_{\rm max} \sim 0.03$ and $0.2 \ h \ {\rm Mpc}^{-1}$
the 68\% MCIs do not quite touch each other. However, this discrepancy
is not statistically significant. We bring up this point here
nonetheless as a cautionary note against over-interpretion of credible
intervals.

Nonlinear correction also
helps to improve consistency in the $\Omega_m h$ estimates from 2dF in
figure~\ref{fig:2dF}. As for SDSS-
2~main, the large uncertainties in
$\Omega_m h$ in figure~\ref{fig:main} means that internal consistency
is not an issue, with or without nonlinear correction.

\paragraph{How much correction?}
Focussing on the cases with nonlinear correction, we see in
figures~\ref{fig:LRG} to \ref{fig:main}
that while SDSS-4~LRG
clearly prefers a nonzero correction
parameter---be it $P_{\rm shot}$ or $Q_{\rm nl}$---at $k_{\rm max} \gwig 0.1 \ h \
{\rm Mpc}^{-1}$,
the 2dF and the SDSS-2~main data sets do not exhibit the same strong preference.
For SDSS-2~main and most choices of $k_{\rm max}$ for 2dF, the $1 \sigma$
regions include $P_{\rm shot}=0$ and
$Q_{\rm nl}=0$ (although,
as mentioned before, the case of $Q_{\rm nl}=0$ is not  equivalent to
no nonlinear correction).  At $k_{\rm max} \lwig 0.1 \ h \
{\rm Mpc}^{-1}$, the parameters $P_{\rm shot}$ and $Q_{\rm nl}$ cannot
be constrained by data.

In terms of the $P$~model, roughly half of the small scale ($k \gwig 0.1 \ h \ {\rm Mpc}^{-1}$)
power in SDSS-4~LRG can be attributed to the shot noise term.  For 2dF and SDSS-2~main, the
shot noise contribution is about a quarter according to the 1D mode.

\paragraph{Which nonlinear model?}
Perhaps the most striking feature about the two nonlinear models considered here is
that their corrective effects appear to be identical.  This is
particularly apparent in figures~\ref{fig:LRG} and \ref{fig:2dF} at
$k_{\rm max} \gwig 0.1 \ h \ {\rm Mpc}^{-1}$,
where the preferred values of the
correction parameters $P_{\rm shot}$ and $Q_{\rm nl}$
exhibit virtually the same dependence on $k_{\rm max}$.
At $k_{\rm max} \sim 0.2 \ h \ {\rm Mpc}^{-1}$ the $\Omega_m h$
estimates also show little if any dependence on the priors imposed on
the nonlinear correction parameters.
Thus as far as vanilla cosmology is concerned, there is
no preference for either nonlinear model from a phenomenological standpoint, although
the transparency of the $P$~model still makes it the more attractive one of the two.

\subsection{The external test: are all data sets consistent with each other?}

\begin{figure}[t]
\hspace{25mm}
\includegraphics[width=10cm]{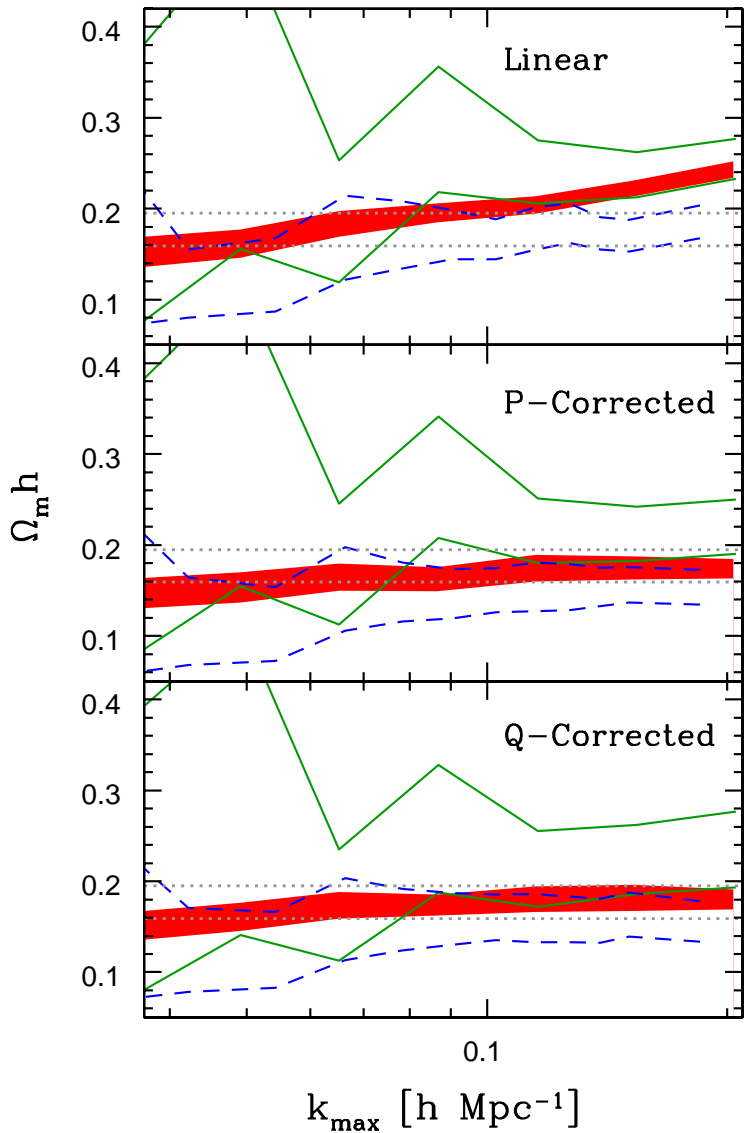}
\caption{1D marginal 68\% MCIs for $\Omega_m h$ as functions of $k_{\rm max}$ using
galaxy clustering data from SDSS-4 LRG (red/shaded region), 2dF
(blue/dashed line) and SDSS-2 main (green/solid line).  {\it Top}: No
nonlinear correction is used.  {\it Middle}: Correction with the
$P$~model.  {\it Bottom}: Correction with the $Q$~model.  For
comparison we show also the corresponding 68\% MCI from WMAP-3
(grey/dotted line).\label{fig:omegamh} }
\end{figure}

Figure~\ref{fig:omegamh} shows how the constraints on $\Omega_m h$
obtained from different galaxy clustering data sets and with different
nonlinear correction methods compare with each other.  For good
comparison we indicate in the figure also the corresponding estimate
from \mbox{WMAP-3}.  Similar information is available in
table~\ref{tab:intervals}, in which we give the 1D marginal 68\% and
95\% MCIs at $k_{\rm max}\sim 0.2 \ h \ {\rm Mpc}^{-1}$ ($k_{\rm max}
\sim 0.18 \ h \ {\rm Mpc}^{-1}$ for 2dF).

%%%%%%%%%%%%%%%%%%%%%%%%%%%%%%%%%%%%%%%%%%%%%%%%%%
%%%%%%%%%%%%%%%%%%%%%%%%%%%%%%%%%%%%%%%%%%%%%%%%%%
%%%%%%%%%%%%%%%%%%%%%%%%%%%%%%%%%%%%%%%%%%%%%%%%%%
%%%%%%%%%%%%%%%%%%%%%%%%%%%%%%%%%%%%%%%%%%%%%%%%%%

%%%%%%%%%%%%%%%%%%%%%%%%%%%%%%%%%%%%%%%%%%%%%%%
%%%%%%%%%%%%%%%%%%%%%%%%%%%%%%%%%%%%%%%%%%%%%%%%
\begin{table}
{\footnotesize
\caption{1D marginal 68\% (95\%) MCIs for $\Omega_m h$ and the nonlinear parameters
$P_{\rm shot}$ and $Q_{\rm nl}$ from the three galaxy clustering data
sets under consideration.  We show results for $k_{\rm max} \sim 0.2 \
h \ {\rm Mpc}^{-1}$ (\mbox{$k_{\rm max} \sim 0.18 \ h \ {\rm
Mpc}^{-1}$} for 2dF), using (i)~no nonlinear correction,
(ii)~correction with the $P$~model, and (iii)~correction with the
$Q$~model.  The WMAP-3 preferred regions are also quoted here for
comparison.
\label{tab:intervals}}
\begin{tabular}{lccc}
 \br Data set & No correction & $P$ model & $Q$ model \\ \mr $\Omega_m
 h$ \\
 SDSS-4 LRG & 0.234--0.252 (0.227--0.262) & 0.163--0.184 (0.153--0.195) & 0.169--0.191 (0.158--0.202) \\
 2dF & 0.167--0.204(0.152--0.228) & 0.134--0.173 (0.122--0.196) & 0.134--0.178 (0.118--0.207) \\
 SDSS-2 main & 0.233--0.277 (0.213--0.300) & 0.190--0.250 (0.167--0.279) & 0.193--0.276 (0.167--0.316) \\
 WMAP-3 & & 0.159--0.195 (0.142--0.211) \\
\mr
$P_{\rm shot}$ or $Q_{\rm nl}$\\
SDSS-4 LRG & -- & 3980--5170 (3380--5740) & 22.7--25.5 (18.6--37.9) \\
2dF & -- & 452--1310 ($<1640$) & 7.0--19.2 (2.5--27.3) \\
SDSS-2 main & -- & $< 506$ ($< 831$) & $<10.2$ ($<19.5$) \\
\br
\end{tabular}
}
\end{table}
%%%%%%%%%%%%%%%%%%%%%%%%%%%%%%%%%%%%%%%%%%%%%%%%%%%
%%%%%%%%%%%%%%%%%%%%%%%%%%%%%%%%%%%%%%%%%%%%%%%%%%%

For both SDSS-4 LRG and SDSS-2 main, nonlinear correction clearly
leads to better agreement with 2dF and with WMAP-3 at $k_{\rm max}
\gwig 0.1 \ h \ {\rm Mpc}^{-1}$.  In the case of \mbox{SDSS-2 main},
although the correction is not sufficient to cause the 68\% region to
overlap with that from 2dF (using the same correction method), the two
95\% regions are certainly consistent.  Importantly, the level of
disagreement between \mbox{SDSS-2 main} and 2dF after correction is no
worse than the small internal inconsistency between the low and the
high $k_{\rm max}$ constraints on $\Omega_m h$ from SDSS-4 LRG already
discussed in section~\ref{sec:results}.  Thus if we should accept that
the nonlinear models~(\ref{eq:Qmodel}) and (\ref{eq:Pmodel}) offer
sufficient correction for SDSS-4 LRG, logically we must also consider
them adequate for \mbox{SDSS-2 main}.

Lastly, we observe in figure~\ref{fig:omegamh} that the 2dF preferred
values of $\Omega_m h$ tend in any case to be on the low side of
SDSS-2 main, even at values of $k_{\rm max}$ well below those at which
nonlinearity nominally sets in. This suggests that the residual
inconsistency after nonlinear correction can rather be put down to a
statistical aberration at small $k$ values, than is indicative of a
failure of either nonlinear model.  Indeed, Sanchez and
Cole~\cite{Sanchez:2007rc} compared directly the power spectra of red
galaxies (believed to be the main source of scale-dependent biasing)
from 2dF and the main galaxy sample of SDSS data release~5, and found
a similar discrepancy in the raw power spectrum data 
(see figure~7 of their paper).  Our
\mbox{SDSS-2 main} data set is but a subsample of SDSS data release~5;
that it contains the same small fluctuation should be of no surprise.

\section{Test 2: vanilla+massive neutrinos \label{sec:neutrinos}}

It is well known that a subdominant component of massive neutrino HDM
in the matter content slows down the growth of density perturbations
on small length scales.  In terms of the matter power spectrum, we
expect a suppression of power of order $\Delta P_{\rm lin}/P_{\rm lin}
\sim 8 \ \Omega_\nu/\Omega_m$ at $k \gg k_{\rm FS}$, where $k_{\rm
FS}$ is the neutrinos' free-streaming wavenumber, and $\Omega_\nu$ the
neutrino energy density. Since this suppression effectively changes
the shape of the matter power spectrum at large wavenumbers, some
amount of degeneracy could conceivably exist between the neutrino
energy density and the nonlinear correction parameters.  In this
section we investigate the possible existence of such a degeneracy,
and if so, its effects on massive neutrino cosmology.

\subsection{Set-up\label{sec:fnu-setup}}

We consider three parameter spaces:
\begin{enumerate}
\item $f_\nu, \Omega_m h,\Omega_b h^2,h,n_s,\ln (10^{10} {\cal A})$,
\item $f_\nu, \Omega_m h, \Omega_b h^2,h,n_s,\ln (10^{10} {\cal A} ),
P_{\rm shot}$, and
\item $f_\nu, \Omega_m h,\Omega_b h^2,h,n_s,\ln (10^{10} {\cal A}), Q_{\rm nl}$.
\end{enumerate}
Here, the neutrino fraction $f_\nu \equiv \Omega_\nu/\Omega_m$ is
defined as the ratio of the neutrino energy density~$\Omega_\nu$ to the total
matter density $\Omega_m$.  The former is given by the well known
expression
\begin{equation}
\Omega_\nu h^2 =  \frac{\sum m_\nu}{93 \ {\rm eV}},
\end{equation}
where $\sum m_\nu$ denotes the sum of the neutrino masses.  We assume
3.04 degenerate neutrino species, which should be a good approximation
since neither CMB nor galaxy clustering measurements are at present
sufficiently sensitive to $\sum m_\nu \lwig 0.3 \ {\rm eV}$, where one
might expect some small effects due to the neutrino mass hierarchy.

We fit these three cases to SDSS-4 LRG up to $k_{\rm max} \sim 0.2 \ h
\ {\rm Mpc}^{-1}$, using no nonlinear correction in case~(i), and
correction with the $P$ and the $Q$~model respectively for cases~(ii)
and (iii).  As in the previous section, we impose WMAP-3 priors on the
parameters $h$, $n_s$, and $\Omega_b h^2$ tabulated in
table~\ref{tab:priors}.  Note that these priors differ from those used
previously, since CMB constraints are model-dependent.  We adopt a
top-hat prior on $f_\nu$, 0--0.5, while for $Q_{\rm nl}$ we use
0--100, in anticipation that more nonlinear correction may be required
to offset the higher $\Omega_m h$ values usually inferred in
cosmologies with massive neutrinos.

\subsection{Results and discussions}

Table~\ref{tab:nu} shows the 1D 68\% and 95\% MCIs for $\Omega_m h$,
$f_\nu$, and the nonlinear parameters $P_{\rm shot}$ and $Q_{\rm nl}$
for the three cases considered.  We also give the minimum $\chi^2$
values as a measure of the goodness-of-fit.

%%%%%%%%%%%%%%%%%%%%%%%%%%%%%%%%%%%%%%%%%%%%%%%%%%%%%%%%%%%%%%%%%%%%%%%%%%%%%%
%%%%%%%%%%%%%%%%%%%%%%%%%%%%%%%%%%%%%%%%%%%%%%%%%%%%%%%%%%%%%%%%%%%%%%%%%%%%%%
\begin{table}
{\footnotesize
\caption{1D marginal 68\% (95\%) MCIs for $\Omega_m h$, $f_\nu$, and
the nonlinear parameters $P_{\rm shot}$ and $Q_{\rm nl}$ from SDSS-4
LRG, and the associated minimum $\chi^2$ values.  These are obtained
using (i)~no nonlinear correction, (ii)~correction with the $P$~model,
and (iii)~correction with the $Q$~model. We quote here also the WMAP-3
preferred regions for comparison.\label{tab:nu}}
\begin{tabular}{lcccc}
\br
Model & $\Omega_m h$ & $f_\nu$ & $P_{\rm shot}$ or $Q_{\rm nl}$ & $\chi^2_{\rm min}$\\
\mr
No correction & 0.246--0.287 (0.232--0.314) & 0.01--0.047 ($<0.065$) & -- & 63.2 \\
$P$~model & 0.168--0.216 (0.154--0.256)  & $<0.057$ ($<0.112$) & 4070--5270 (3430--5810) & 19.7 \\
$Q$~model  & 0.171--0.226  (0.158--0.274) & $<0.061$ ($<0.135$) & 23.6--36.9 (17.5--47.5) & 20.6\\
\mr
WMAP-3 & 0.178--0.245 (0.151--0.270) & $<0.080$ ($<0.127$) & -- & -- \\
\br
\end{tabular}
}
\end{table}
%%%%%%%%%%%%%%%%%%%%%%%%%%%%%%%%%%%%%%%%%%%%%%%%%%%%%%%%%%%%%%%%%%%%%%%%%%%%%
%%%%%%%%%%%%%%%%%%%%%%%%%%%%%%%%%%%%%%%%%%%%%%%%%%%%%%%%%%%%%%%%%%%%%%%%%%%%%

As in the case for vanilla cosmology, fitting the SDSS-4 LRG data
without nonlinear correction leads to a very poor goodness-of-fit; the
$\chi_{\rm min}^2$ value is 63.2, for approximately $20+3-6=17$
degrees of freedom (20 data points
of SDSS-4 LRG, 3 for the priors on $n_s$, $h$ and $\Omega_b h^2$, and $-6$
for 6 free parameters).  Adding nonlinear correction reduces
$\chi^2_{\rm min}$ by more than 40 units at the expense of only one
extra parameter. Again, both nonlinear models offer very similar
corrections to the power spectrum in terms of the inferred $\Omega_m
h$ and $f_\nu$ values, although the $P$~model appears to provide a
slightly better fit to the data judging by its slightly smaller
$\chi^2_{\rm min}$.  Interestingly, despite the dramatic decrease in
the minimum $\chi^2$ between correction and no correction, the
resulting shifts in the $\Omega_m h$ and$f_\nu$ estimates are
deceptively small so that the 95\% MCIs remain compatible before and
after correction.

In the absence of nonlinear correction, the 95\% upper limit on
$f_\nu$ is about a factor of two too tight compared with the corrected
case.  This situation is reminiscent of the overly constraining bounds
on $\sum m_\nu$ derived in some recent combined analyses of WMAP-3 and
the flux power spectrum of the Lyman-$\alpha$ forest~\cite{Seljak:2006bg}.
Too much power
at large wavenumbers---either because of uncorrected nonlinearities in
the galaxy power spectrum or an unusually large normalisation in the
case of the Lyman-$\alpha$ forest---leads to the appearance of an
overly flat matter power spectrum, which in turn prefers a smaller
neutrino fraction and hence a smaller neutrino mass.  The difference
between the two corrected $f_\nu$ bounds is about 20\%.  The
corresponding 95\% limits on $\sum m_\nu$, $<1.76 \ {\rm eV}$ for the
$P$~model and $<1.83 \ {\rm eV}$ for the $Q$~model, differ by even
less. Thus cosmological neutrino mass determination is at present
unaffected by our choice of nonlinear correction model.

%%%%%%%%%%%%%%%%%%%%%%%%%%%%%%%%%%%%%%%%%%%%%%%%%%%%%%%%%%%%%
%%%%%%%%%%%%%%%%%%%%%%%%%%%%%%%%%%%%%%%%%%%%%%%%%%%%%%%%%%%%%
\begin{figure}[t]
%\hspace{10mm}
\includegraphics[width=6.2cm,angle=270]{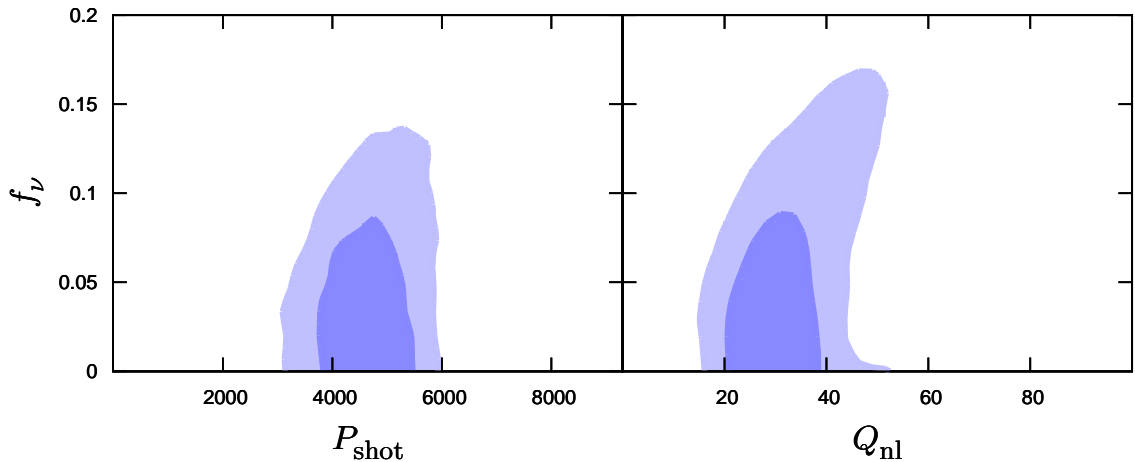}
\caption{2D marginal 68\% and 95\% MCIs for $\{f_\nu,P_{\rm shot}\}$
and $\{f_\nu, Q_{\rm nl}\}$ from \mbox{SDSS-4 LRG}, 
for cases (ii) and (iii)
of section~\ref{sec:fnu-setup}.\label{fig:fnu}}
\end{figure}
%%%%%%%%%%%%%%%%%%%%%%%%%%%%%%%%%%%%%%%%%%%%%%%%%%%%%%%%%%%
%%%%%%%%%%%%%%%%%%%%%%%%%%%%%%%%%%%%%%%%%%%%%%%%%%%%%%%%%%%

Finally, we see in figure~\ref{fig:fnu} that no strong degeneracy
exists between the neutrino fraction $f_\nu$ and the nonlinear
correction parameters $P_{\rm shot}$ and $Q_{\rm nl}$.  For small
values of~$f_\nu$, the changes induced in the linear matter power
spectrum by massive neutrino dark matter can be approximately mimicked
by a redefinition of the apparent $\Omega_m h$
parameter~\cite{Cole:2005sx},
\begin{equation}
(\Omega_m h)_{\rm apparent} = (\Omega_m h)_{\rm true} - 1.2 f_\nu,
\end{equation}
assuming $h \sim 0.7$.  Here, $(\Omega_m h)_{\rm apparent}$ is the
parameter that is actually constrained by power spectrum data, while
$(\Omega_m h)_{\rm true}$ denotes the true value of $\Omega_m h$. This
expression also encapsulates the well known (approximate) degeneracy
between $\Omega_m h$ and $f_\nu$.  Larger $f_\nu$ values induce more
complicated changes in the power spectrum, and it is reassuring to see
that these changes are not degenerate with nonlinear correction using
either model.

\section{The pathology of the $Q$ model: vanilla+thermal axions \label{sec:axions}}

While both nonlinear models work very well for vanilla and for
vanilla+massive neutrino cosmologies, and we may be tempted to
conclude that they are for all purposes phenomenologically identical,
we provide in this section a counter-example in which the incorrect
functional form of the $Q$~model can lead to some misleading results.

The case in point is a class of models containing a possible
subdominant HDM component due to relic thermal axions with mass $m_a$.
These models differ from those with massive neutrino HDM in that the
temperature $T_a = [10.75/g^*_D(m_a)]^{1/3} \ T_\nu$ and hence the
number density $n_a = [\zeta(3)/\pi^2] \ T_a^3$ of the axions are
functions of $m_a$, since $m_a$ determines when the particle species
should decouple from the primordial plasma.  Here, the function
$g^*_D(m_a)$ denotes the effective number of thermal degrees of
freedom at the time of decoupling, and must be calculated by carefully
tracking the freeze-out process (e.g., \cite{Hannestad:2005df}).

Thus, although qualitatively thermal axion HDM exhibits free-streaming
features very similar to those of massive neutrinos, quantitatively
the suppression of small scale power in the matter power spectrum has
a nonlinear dependence on the axion mass.   We show in
this section how this nontrivial dependence can cause some problems
for the $Q$~model.

\subsection{Set-up \label{sec:axion-setup}}

We consider three parameter spaces:
\begin{enumerate}
\item $m_a, \Omega_m h,\Omega_b h^2,h,n_s,\ln (10^{10} A_s), b, Q_{\rm
nl}$, with a top-hat prior 0--100 on $Q_{\rm nl}$,
\item $m_a, \Omega_m h,\Omega_b h^2,h,n_s,\ln (10^{10} A_s), b, Q_{\rm
nl}$, with a top-hat prior 0--200 on $Q_{\rm nl}$, and
\item $m_a, \Omega_m h, \Omega_b h^2,h,n_s,\ln (10^{10} A_s), b, P_{\rm shot}$.
\end{enumerate}
Cases (i) and (ii) both use the nonlinear model~(\ref{eq:Qmodel}), and
differ only in the priors imposed on the correction parameter $Q_{\rm
nl}$.  Case~(iii) uses the nonlinear model~(\ref{eq:Pmodel}), with the
usual prior on $P_{\rm shot}$ (see table~\ref{tab:priors}).  We fit
each case to the combined data set WMAP-3+SDSS-4 LRG, using data up to
$k_{\rm max} \sim 0.2 \ h \ {\rm Mpc}^{-1}$ for the latter.

\subsection{Results and discussions}

\begin{figure}[t]
%\hspace{25mm}
\includegraphics[width=10.8cm,angle=270]{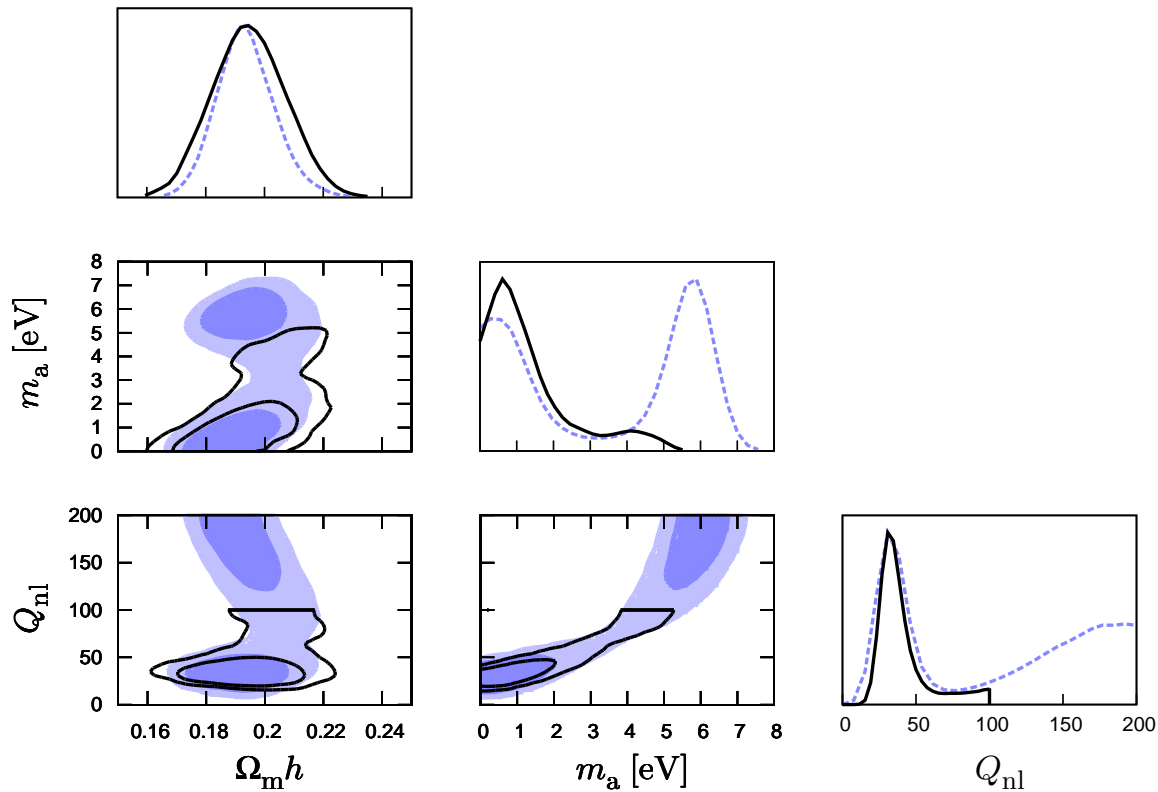}
\caption{2D marginal 68\% and 95\% MCIs for $\{\Omega_m h,m_a\}$,
$\{\Omega_m h, Q_{\rm nl}\}$, and $\{m_a,Q_{\rm nl}\}$ from
WMAP-3+SDSS-4 LRG.  In the diagonal are the 1D marginal posteriors for
the three named parameters.  Black/solid lines correspond to case (i)
of section~\ref{sec:axion-setup} with a top-hat prior 0--100 on
$Q_{\rm nl}$, while the blue/shaded regions and blue/dashed lines
denote case (ii) with $Q_{\rm nl}$ prior 0--200.\label{fig:axionQ}}
\end{figure}

Figure~\ref{fig:axionQ} shows the 2D marginal 68\% and 95\% MCIs for
$\{\Omega_m h,m_a\}$, $\{\Omega_m h, Q_{\rm nl}\}$,
and $\{m_a,Q_{\rm nl}\}$, as well as the 1D marginal
posteriors for the the same three parameters for cases (i) and (ii).

%%%%%%%%%%%%%%%%%%%%%%%%%%%%%%%%%%%%%%%%%%%%%%%%%%%%%%%%
%%%%%%%%%%%%%%%%%%%%%%%%%%%%%%%%%%%%%%%%%%%%%%%%%%%%%%%%
%%%%%%%%%%%%%%%%%%%%%%%%%%%%%%%%%%%%%%%%%%%%%%%%%%%
%%%%%%%%%%%%%%%%%%%%%%%%%%%%%%%%%%%%%%%%%%%%%%%%%%%

Consider first the 1D marginal posteriors for $Q_{\rm nl}$.  We see in
figure~\ref{fig:axionQ} that the posteriors in both cases (i) and (ii) remain
finite all the way up to the upper limit of the prior imposed on
$Q_{\rm nl}$.  This is also reflected in the relevant 2D contours,
which are abruptly cut off at $Q_{\rm nl} = 100$ and $Q_{\rm nl}=200$
respectively.  Data alone does not constrain $Q_{\rm nl}$ in this
class of cosmological models.

%%%%%%%%%%%%%%%%%%%%%%%%%%%%%%%%%%%%%%%%%%%%%%%%%%%%%%%%
%%%%%%%%%%%%%%%%%%%%%%%%%%%%%%%%%%%%%%%%%%%%%%%%%%%%%%%%
\begin{figure}[t]
\hspace{25mm}
\includegraphics[width=8.5cm]{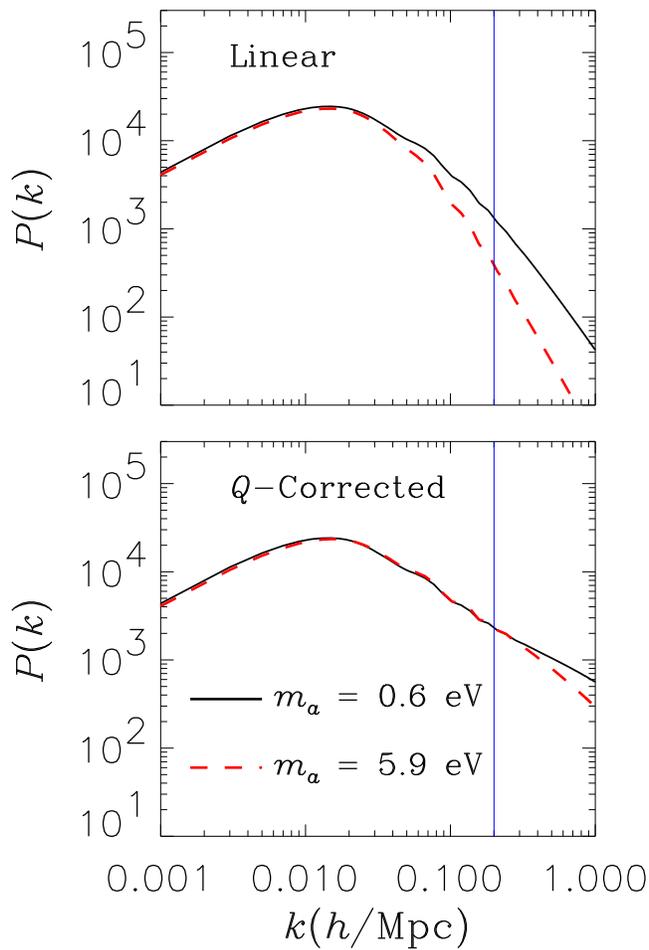}
\caption{Matter power spectra for the two peak values of $m_a$. 
The black/solid lines are for $\{m_a = 0.6 \ {\rm eV}, Q_{\rm nl} = 31\}$, and the red/dashed lines
for $\{m_a =5.9 \ {\rm eV}, Q_{\rm nl} = 170\}$. 
The upper panel shows the linear matter power spectra, while 
the lower panel includes nonlinear correction with the $Q$~model.  The vertical line
indicates the maximum value of $k$ used in the analyses.\label{fig:mpk}}
\end{figure}
%%%%%%%%%%%%%%%%%%%%%%%%%%%%%%%%%%%%%%%%%%%%%%%%%%%
%%%%%%%%%%%%%%%%%%%%%%%%%%%%%%%%%%%%%%%%%%%%%%%%%%%

While this does not affect the $\Omega_m h$ estimates, the inference
of the axion mass $m_a$ depends crucially on how well we can constrain
$Q_{\rm nl}$ because of a persisting degeneracy between the two
parameters.  Indeed, if we do not cut off $Q_{\rm nl}$ by hand at some
sufficiently small value, a second peak begins to appear in the
posterior at \mbox{$\{m_a \sim 6 \ {\rm eV}, Q_{\rm nl} \sim 200\}$},
besides the one at
\mbox{$\{m_a \sim 0 \ {\rm eV},Q_{\rm nl} \sim 30\}$}. 
From figure~\ref{fig:mpk} we see that 
the power spectra for the two peaks after nonlinear correction
are almost perfectly degenerate up to (and beyond) 
$k \sim 0.2 \ h \ {\rm Mpc}^{-1}$, 
even though the corresponding linear matter power spectra
differ markedly already at $k \sim 0.04 \ h \ {\rm Mpc}^{-1}$.

We may be inclined to regard a 6~eV axion and, by implication, a large $Q_{\rm nl}$ value
as unphysical because the
former runs in conflict with constraints on $m_a$ derived from
stellar energy loss arguments and from telescope searches for axion
radiative decays~\cite{Raffelt:2006cw}.
Furthermore, the $m_a,Q_{\rm nl}$-degeneracy exists only in a limited $k$ range:
as shown in figure~\ref{fig:mpk}, 
the corrected power spectra for the two peaks begin to deviate at $k \gwig 0.3 \ h \ {\rm Mpc}^{-1}$. 
This suggests that the exact location of the  high~$m_a$ peak in $\{m_a,Q_{\rm nl}\}$-space 
may in fact depend on our choice of $k_{\rm max}$.
Naturally we could have avoided this second peak simply by demanding 
consistency with other astrophysical constraints on~$m_a$.
However, 
in the absence of, e.g., galaxy formation simulations for this very
class of cosmological models, we have {\it a priori} no
reason to reject $Q_{\rm nl} \sim 200$ or any other higher or lower value
besides our own
prejudices.  Moreover, even if we manage to avoid the second peak
with a finely tuned prior on $Q_{\rm nl}$, the $m_a,Q_{\rm
nl}$-degeneracy means that the constraint thus obtained on $m_a$ will
still depend sensitively on exactly what we choose to be the upper
limit of that prior.

 %%%%%%%%%%%%%%%%%%%%%%%%%%%%%%%%%%%%%%%%%%%%%%%%%%%%%%%%%%%%%
%%%%%%%%%%%%%%%%%%%%%%%%%%%%%%%%%%%%%%%%%%%%%%%%%%%%%%%%%%%%%
\begin{figure}[t]
%\hspace{25mm}
\includegraphics[width=10.8cm,angle=270]{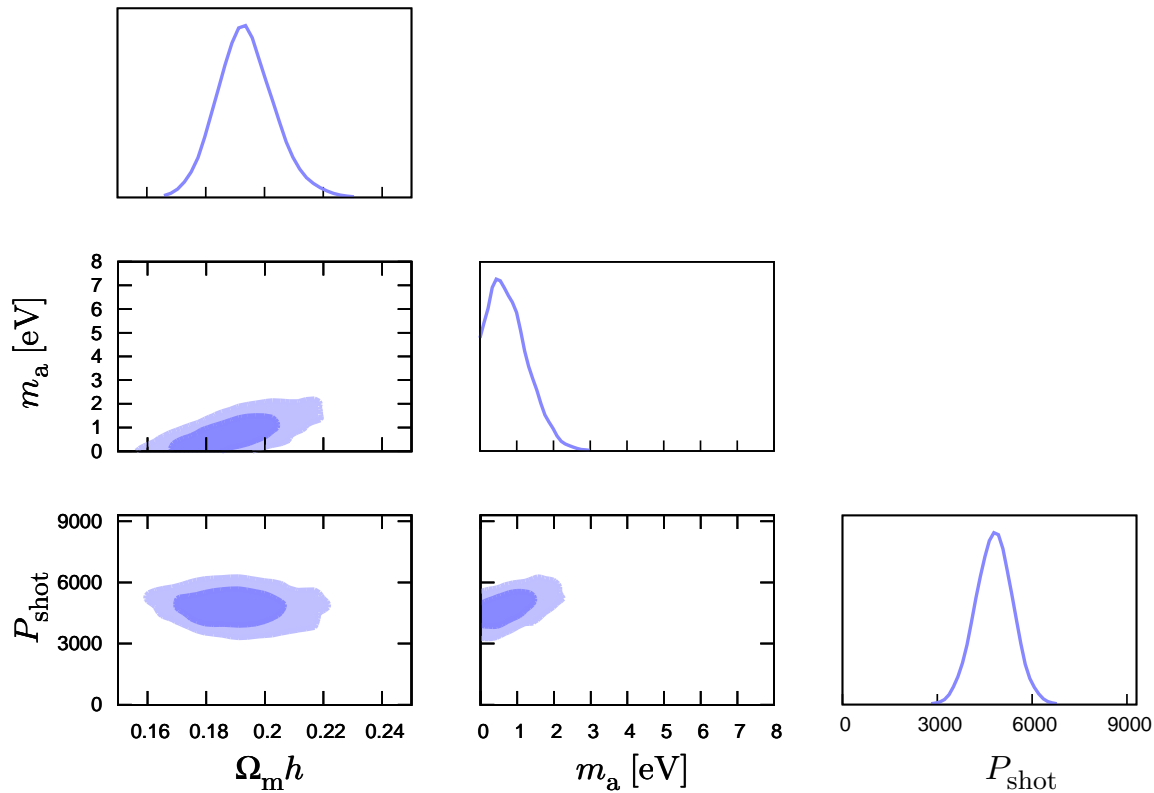}
\caption{2D marginal 68\% and 95\% MCIs for $\{\Omega_m h,m_a\}$, $\{\Omega_m h, P_{\rm shot}\}$,
and $\{m_a,P_{\rm shot}\}$ from WMAP-3+SDSS-4 LRG for case (iii) of section~\ref{sec:axion-setup}.
1D marginal posteriors for the same three parameters are shown in the diagonal.\label{fig:axionP}}
\end{figure}
%%%%%%%%%%%%%%%%%%%%%%%%%%%%%%%%%%%%%%%%%%%%%%%%%%%%%%%%%%%
%%%%%%%%%%%%%%%%%%%%%%%%%%%%%%%%%%%%%%%%%%%%%%%%%%%%%%%%%%%

This exercise also  highlights the danger of analytic
marginalisation~\cite{Lewis:2002ah},
a technique sometimes used on nuisance parameters
in {\sc CosmoMC} to shorten the computation time.
Here, analytic integration of the posterior in the direction of
a nuisance parameter is made possible
by taking the lower and upper limits of the parameter's top-hat prior to
$-\infty$ and $\infty$ respectively.  Using this technique on the marginalisation of
$Q_{\rm nl}$, we find a unimodal 1D posterior for $m_a$ whose
 95\% MCI of $4.95 \lwig  m_a/{\rm eV} \lwig 7.17$
favours unambiguously the high~$m_a$ region.
Thus, analytic marginalisation can
be very useful if the likelihood function itself sufficiently
constrains the nuisance parameters.  Otherwise, as we have seen here,
the end results are potentially misleading.

Finally, we note that the $P$~model does not suffer these problems.  Figure~\ref{fig:axionP}
shows the 2D marginal 68\% and 95\% MCIs for
$\{\Omega_m h,m_a\}$, $\{\Omega_m h, P_{\rm shot}\}$,
and $\{m_a,P_{\rm shot}\}$, and the corresponding 1D marginal posteriors for
case (iii).  Here, although the correction parameter~$P_{\rm shot}$ is slightly degenerate
with the axion mass $m_a$, it is independently well constrained by data.
Importantly, even if data fails to constrain $P_{\rm shot}$,
we have a simple and well defined way to choose our prior on $P_{\rm shot}$.  Thus we
conclude that the $P$~model is at present superior to the $Q$~model for nonlinear correction in
non-vanilla cosmologies.

\section{Conclusions \label{sec:conclusions}}

In this paper we have explored two nonlinear correction and scale-dependent bias
models---the $Q$~model of reference~\cite{Cole:2005sx} and the halo model-inspired
$P$~model---in some detail.  We have confronted them with a range of galaxy clustering
data and cosmologies to determine the strengths and weaknesses of the models.

In the context of standard $\Lambda$CDM cosmology, we find that both models
perform equally well on present galaxy power spectrum data, in the sense that their
corrective effects on the matter power spectrum are essentially identical.
The case for nonlinear correction is very strong for the SDSS-4 LRG power
spectrum.  An indicative figure of merit is the minimum $\chi^2$: fitting data up to
$k_{\rm max} \sim 0.2 \ h \ {\rm Mpc}^{-1}$, we find that $\chi^2_{\rm min}$ changes
from $\sim 60$ for 18 degrees of freedom
without correction to $\sim 20$ for 17 degrees of freedom with correction, irrespective of
the nonlinear model used.
For 2dF and \mbox{SDSS-2 main}, however, the need for correction is marginal
and subsists primarily because their respective $\Omega_m h$ estimates show better
agreement with than without nonlinear correction.  The preferred
values of $\Omega_m h$ from
SDSS-4 LRG, SDSS-2 main, and 2dF after correction, as well as from WMAP-3
can all be reconciled at 95\% confidence,
contrary to the case without correction.

Similar results are obtained for $\Lambda$CDM cosmologies extended with a subdominant
component of massive neutrino hot dark matter.  Data again show no strong
preference for either nonlinear correction model, nor do we find any detrimental
degeneracy between the neutrino fraction $f_\nu$ and either nonlinear correction
parameter.  Cosmological neutrino mass determination is at the time being
unaffected by our choice
of nonlinear correction model.

However, if the subdominant free-streaming dark matter is in the form of relic thermal
axions, a nontrivial degeneracy between the axion mass $m_a$ and the correction parameter
$Q_{\rm nl}$ renders the $Q$~model highly pathological, so that our inference of $m_a$
depends sensitively on the prior we impose on $Q_{\rm nl}$.  In contrast, the $P$~model,
whose functional form is based on well motivated physics,
does not suffer from this problem, and is arguably superior to the $Q$~model.
Note that we have used relic thermal axions here as an example.
But our results may also be relevant for other light thermal relics whose temperature and
hence abundance depend on the mass of the particle.

More precise data from future galaxy redshift surveys will eventually render these
simplistic models inadequate to describe nonlinear evolution and scale-dependent biasing.
For example, the damping of baryon acoustic oscillations due to nonlinear mode-coupling
will need to be factored into the game at some stage~\cite{Smith:2006ne,Crocce:2007dt,Smith:2007gi}.
The advent of wide- and deep-field
weak gravitational lensing surveys in the next decade as an alternative probe of the large scale structure
distribution will circumvent some of these nonlinearity issues.  But
galaxy redshift surveys will remain an important tool for the observation of
baryon acoustic oscillations, and nonlinear evolution/scale-dependent bias modelling will continue
to constitute an important aspect of the cosmological analysis machinery.

\section*{Acknowledgements}

We thank Alexia Schulz and Robert E.~Smith for useful comments on the manuscript.
We acknowledge use of computing resources from the Danish Center
for Scientific Computing (DCSC).

\end{document}